\documentclass[12pt]{article}
\usepackage{amsfonts,amstext,latexsym,times,xspace}
\usepackage{epsfig}

\newcommand{\lb}[1]{\label{#1}}

\renewcommand{\[}{\begin{eqnarray}}
\renewcommand{\]}{\end{eqnarray}}
\newcommand{\nn}{\nonumber}
\newcommand{\non}{\nonumber \\ }
\renewcommand{\=}{\equiv}
\def\ba{\begin{array}}
\def\ea{\end{array}}
\makeatletter
\newif\if@fewtab\@fewtabtrue
\def\moth{\mathsurround=0pt}
\newdimen\zo \zo=0pt

\def\tick{\leaders\hrule height 0.5ex depth 0pt \hskip 0.5pt}
\def\upboxfill{$\moth \setbox\zo\hbox{\tick}%
  \hskip 2pt\hbox to 0pt{$\tick$\hss}\hrulefill \hbox to 2pt{$\tick$\hss}$}
\def\underbox#1{\offinterlineskip{\mathord{\mathop{\vtop{\moth\ialign{##\crcr
      $\hfil\displaystyle{#1}\hfil$\crcr\noalign{}
      {\upboxfill}\crcr\noalign{}}}}\limits}}}
\def\dtick{\leaders\hrule height .34pt depth 0.5ex \hskip 0.5pt}
\def\downboxfill{$\moth \setbox\zo\hbox{\dtick}%
  \hskip 2pt\hbox to 0pt{$\dtick$\hss}\hrulefill%
  \hbox to 2pt{$\dtick$\hss}$}
\def\overbox#1{\mathop{\vbox{\moth\ialign{##\crcr\noalign{}
\downboxfill\crcr\noalign{\vskip 1pt\nointerlineskip}
      $\hfil\displaystyle{#1}\hfil$\crcr}}}\limits}

\newcommand{\undersym}[1]{\underbox{{}#1}}
\newcommand{\oversym}[1]{\!\overbox{{}#1}}

\newcommand{\cD}{\ensuremath{\mathcal{D}}\xspace}

\newcommand{\6}[1][{}]{\ensuremath{E_{6 #1}}\xspace}
\newcommand{\7}[1][{}]{\ensuremath{E_{7 #1}}\xspace}

\newcommand{\E}{E_{8(8)}}
\newcommand{\EE}{E_{7(7)}}
\newcommand{\EEE}{E_{6(6)}}

\newcommand{\USp}[1][8]{\ensuremath{\mbox{USp$(#1)$}}\xspace}

\newcommand{\SLR}[1][8]{\ensuremath{\mbox{SL$(#1,\mathbb{R})$}}\xspace}

\newcommand{\fg}{\ensuremath{\mathfrak{g}}\xspace}

\newcommand{\gO}[1][{}]{\Omega^{#1}\xspace}
\newcommand{\gOd}[1][{}]{\Omega_{#1}\xspace}

\newcommand{\gd}[1][{}]{\delta_{#1}{}}

\newcommand{\eps}[1][{}]{\epsilon^{#1}\xspace}
\newcommand{\epsd}[1][{}]{\epsilon_{#1}\xspace}

\newcommand{\Zt}{\tilde{Z}{}}
\newcommand{\Zbt}{\bar{Z}{}}

\newcommand{\Et}{\tilde{E}}
\newcommand{\Ht}{\tilde{H}}
\newcommand{\Ft}{\tilde{F}}
\newcommand{\Gt}{\tilde{G}}
\newcommand{\Xt}{\tilde{X}}
\newcommand{\Yt}{\tilde{Y}}

\newcommand{\ft}[2]{{\textstyle {\frac{#1}{#2}} }}

\newcommand{\st}{^\ast}

\newcommand{\rep}[1]{\ensuremath{\mbox{\mathversion{bold}$\mathbf{#1}$%
                     \mathversion{normal}}}}

\renewcommand{\i}{\mathrm{i}}

\newcommand{\cX}{\mathcal{X}}
\newcommand{\cY}{\mathcal{Y}}

\newcommand{\cN}{\mathcal{N}}
\newcommand{\cI}{\mathcal{I}}

\newcommand{\Com}[2]{[#1\, ,\,#2]}
\newcommand{\Sympl}[2]{\left\langle #1,#2\right\rangle}

\newcommand{\Rn}{\ensuremath{\mathbb{R}}\xspace}

\begin{document}

\begin{titlepage}
\begin{center}
\begin{large}
\textbf{ Unitary Realizations of U-duality Groups as Conformal and
Quasiconformal Groups and Extremal
 Black Holes  of \\  Supergravity Theories }\footnote{
 Invited talk at the XIXth Max Born Symposium on Fundamental Interactions and Twistorlike Methods
 ( Wroclaw, Sept. 28 - Oct. 1, 2004).  To appear in the proceedings of the symposium in the  AIP proceedings series .  }
\vspace{1cm}

 Murat G\"unaydin
\\
\end{large}
\emph{Physics Department,
Penn State University\\
University Park, PA 16802, USA \\
email: murat@phys.psu.edu } \\

\begin{abstract}
{\small
 We review the current status of the construction of unitary representations of U-duality groups
 of supergravity theories in five, four and three dimensions. We focus  mainly on the maximal $N=8$ supergravity
 theories and on the $N=2$ Maxwell-Einstein supergravity (MESGT) theories defined by Jordan algebras of degree
 three in five dimensions and their descendants in four and three dimensions. Entropies of the extremal black hole solutions
 of these theories in five and four dimensions  are given by certain invariants of their U-duality groups.
 The five dimensional U-duality groups admit extensions to spectrum generating generalized conformal groups which
 are isomorphic to the U-duality groups of corresponding four dimensional theories. Similarly, the U-duality
 groups of four dimensional theories admit extensions to spectrum generating quasiconformal groups that are
 isomorphic to the corresponding  U-duality groups in three dimensions. For example, the group $E_{8(8)}$ can be realized as a
quasiconformal group in the 57 dimensional charge-entropy space of
BPS black hole solutions of maximal $N=8$ supergravity in four
dimensions and leaves invariant "lightlike separations" with respect
to a quartic norm. Similarly $E_{7(7)}$ acts as a generalized
conformal group in the 27 dimensional charge space of BPS black hole
solutions in five dimensional $N=8$ supergravity and leaves
invariant  "lightlike separations" with respect to a cubic norm. For
the exceptional $N=2$ Maxwell-Einstein supergravity theory the
corresponding quasiconformal and conformal groups are $E_{8(-24)}$
and $E_{7(-25)}$, respectively. We  outline  the oscillator
 construction of the unitary representations of generalized conformal groups that admit positive energy
 representations, which include the U-duality groups of $N=2$ MESGT's in four dimensions . We conclude with  a review of the minimal unitary realizations of U-duality groups
  that are obtained  by  quantizations of  their quasiconformal actions and discuss in detail the minimal unitary realization of
 $E_{8(8)}$.}

\end{abstract}
\end{center}
\end{titlepage}
\newpage
\section{U-Duality Groups in Supergravity Theories}
\subsection{Noncompact exceptional groups as symmetries of maximally extended supergravity theories in various dimensions}
Eleven dimensional supergravity \cite{crjusc} is the effective low energy theory of strongly coupled phase of
M-theory \cite{witten95} and its toroidal compactifications  yield the maximally extended supergravity theories
in $d$ spacetime dimensions with  global non-compact symmetry groups of type $E_{(11-d)(11-d)}$ \cite{julia}. The
discrete subgroups $E_{(11-d)(11-d)}(\mathbb{Z})$ of these groups are expected to be the symmetries of the
non-perturbative spectra of toroidally compactified  M-theory \cite{huto}. We shall use the term U-duality group
to refer to these discrete subgroups as well as to  the global noncompact symmetry groups of supergravity
theories.

In five dimensions  $E_{6(6)}$ is a symmetry of the Lagrangian of the maximal ( $N=8$) supergravity, under which
the 27 vector fields of the theory transform  irreducibly while the 42 scalar fields   transform nonlinearly and
parameterize the coset space \[ \mathcal{M}_5 = E_{6(6)} / USp(8) \]  On the other hand the $E_{7(7)}$ symmetry
of the maximally extended supergravity in $d=4$ is  an on-shell symmetry group. The 28 vector field strengths  of
this theory together with their "magnetic"  duals ( $\tilde{28}$ ) transform irreducibly in the 56 of $E_{7(7)}$
and 70 scalar fields parameterize the coset space \[ \mathcal{M}_4 = E_{7(7)}/SU(8) \]  In three dimensions all
the propagating bosonic degrees of the maximal $N=16$ supergravity can be dualized to scalar fields which
parameterize the coset space  \[ \mathcal{M}_3 = E_{8(8)}/SO(16) \]

\subsection{U-duality groups in Matter coupled Supergravity Theories}
Non-compact global U-duality groups arise in matter coupled supergravity theories as well. In this talk I will
focus mainly on U-duality groups in $N=2$ Maxwell-Einstein supergravity theories (MESGT) in $d=5$ and the
corresponding theories in four and three dimensions. The MESGT's  describe the coupling of an arbitrary number
$n$ of ( Abelian) vector fields to $N=2$ supergravity and five dimensional theories were constructed in
\cite{GST1}. The bosonic part of the Lagrangian can be written as \cite{GST1}

\begin{eqnarray}\label{Lagrange}
e^{-1}\mathcal{L}_{\rm bosonic}&=& -\frac{1}{2}R -\frac{1}{4}{\stackrel{\circ}{a}}_{IJ}F_{\mu\nu}^{I}
F^{J\mu\nu}-\frac{1}{2}g_{xy}(\partial_{\mu}\varphi^{x}) (\partial^{\mu} \varphi^{y}) \nonumber \\ &&+
 \frac{e^{-1}}{6\sqrt{6}}C_{IJK}\varepsilon^{\mu\nu\rho\sigma\lambda}
 F_{\mu\nu}^{I}F_{\rho\sigma}^{J}A_{\lambda}^{K},
\end{eqnarray}
where  $e$ and $R$ denote the f\"{u}nfbein determinant and the scalar curvature in $d=5$ , respectively.
$F_{\mu\nu}^{I}$ are the field strengths of the Abelian vector fields $A_{\mu}^{I} , \,( I=0,1,2 \cdots , n$)
with $A^0_{\mu}$ denoting the "bare" graviphoton. The metric, $g_{xy}$, of the scalar manifold $\mathcal{M}$  and
the "metric" ${\stackrel{\circ}{a}}_{IJ}$ of the kinetic energy term of the vector fields both depend on the
scalar fields $\varphi^{x}$ ( $x,y,..=1,2,..,n$). The invariance under Abelian gauge transformations of the
vector fields requires the completely symmetric tensor $C_{IJK}$ to be constant. Remarkably, one finds that the
entire $N=2$, $d=5$ MESGT is uniquely determined by the constant tensor $C_{IJK}$ \cite{GST1}. In particular ,
the metrics of the kinetic energy terms of the vector and scalar fields are determined by $C_{IJK}$. More
specifically, consider the cubic polynomial, $\mathcal{V}(h)$,
 in $(n+1)$
real variables $h^{I}$ $(I=0,1,\ldots,n)$ defined by the $C_{IJK}$
\begin{equation}
\mathcal{V}(h):=C_{IJK}h^{I}h^{J}h^{K}\ .
\end{equation}
Using this polynomial as a real " K\"ahler potential" for a metric, $a_{IJ}$,
 in an $ n+1 $ dimensional ambient space  with the coordinates  $h^{I}$:
\begin{equation}\label{aij}
a_{IJ}(h):=-\frac{1}{3}\frac{\partial}{\partial h^{I}} \frac{\partial}{\partial h^{J}} \ln \mathcal{V}(h)\ .
\end{equation}
one finds that the   $n$-dimensional    target space, $\mathcal{M}$, of the scalar fields $\varphi^{x}$ can  be
identified with the hypersurface \cite{GST1}
\begin{equation}
\mathcal{V} (h)=C_{IJK}h^{I}h^{J}h^{K}=1
\end{equation}
in this space. The metric  $g_{xy}$ of the scalar manifold is simply  the pull-back of (\ref{aij}) to
$\mathcal{M}$ and the "metric" ${\stackrel{\circ}{a}}_{IJ}(\varphi)$ of the kinetic energy term of the vector
fields appearing in (\ref{Lagrange}) is given by the componentwise restriction of $a_{IJ}$ to $\mathcal{M}$:
\[
{\stackrel{\circ}{a}}_{IJ}(\varphi)=a_{IJ}|_{\mathcal{V}=1} \ .
\]

The physical requirement of  positivity of kinetic energy  requires that  $g_{xy}$ and
${\stackrel{\circ}{a}}_{IJ}$ be positive definite metrics. This requirement induces constraints on the possible
$C_{IJK}$, and in \cite{GST1} it   was shown that any $C_{IJK}$ that satisfy these constraints can be brought to
the following form
\begin{equation}\label{canbasis}
C_{000}=1,\quad C_{0ij}=-\frac{1}{2}\delta_{ij},\quad  C_{00i}=0,
\end{equation}
with  the remaining coefficients $C_{ijk}$
 ($i,j,k=1,2,\ldots , n$) being  completely arbitrary.
 This basis is referred to as the canonical basis for
$C_{IJK}$.

Denoting the symmetry group of the tensor $C_{IJK}$ as $G$ one finds that  the full symmetry group of  $N=2$
MESGT in $d=5$ is of the form \[ G \times SU(2)_R \]  where $SU(2)_R$ denotes the local R-symmetry group of the
$N=2$ supersymmetry algebra. A MESGT is said to be {\it unified} if all the vector fields, including the
graviphoton, transform in an irreducible representation of a {\it simple}  symmetry group $G$ of the lagrangian.
Of all the $N=2$ MESGT's whose scalar manifolds are symmetric spaces only four are unified \cite{GST1}. More
recently it was  shown that if one relaxes the condition that the scalar manifolds be homogeneous spaces then one
finds three novel infinite families ( plus an additional sporadic one ) of unified MESGT's in $d=5$
\cite{gz2003}. If one defines a cubic form \[ \mathcal{N}(h):=C_{IJK}h^{I}h^{J}h^{K} \] using the constant tensor
$C_{IJK}$ , one finds   that the cubic forms associated with the four unified MESGT's can be identified with the
norm forms of simple ( Euclidean) Jordan algebras of degree three \cite{GST1}. There exist only four simple (
Euclidean) Jordan algebras of degree three and they can be realized in terms of $3\times 3$ hermitian matrices
over the four division algebras with the product being one-half the anticommutator. They are denoted as
$J_3^{\mathbb{A}}$, where $\mathbb{A}$ stands for the underlying division algebra , which can be real numbers
$\mathbb{R}$, complex numbers $\mathbb{C}$, quaternions $\mathbb{H}$ and octonions $\mathbb{O}$. The supergravity
theories defined by them were referred to as magical supergravity theories \cite{GST1} since their symmetry
groups in five , four and three dimensions correspond precisely to the symmetry groups of the famous Magic Square
. The octonionic Jordan algebra $J_3^{\mathbb{O}}$ is  the unique exceptional Jordan algebra and consequently the
$N=2$ MESGT defined by it is called the exceptional supergravity theory \cite{GST1}. In the table below we list
the scalar manifolds $G/H$ of the magical supergravity theories in five , four and three dimensions, where $G$ is
the global noncompact symmetry and $H$ is its maximal compact subgroup.

\begin{center}
\begin{small}
\begin{tabular}{|c|c|c|c|}
\hline $J$ & G/H in d=5 & G/H in d=4 & G/H in d=3 \\ \hline
 $J_3^{\mathbb{R}}$ & $SL(3,\mathbb{R})/SO(3)$ & $Sp(6,\mathbb{R})/U(3)$ &$F_{4(4)}/USp(6)\times SU(2)$ \\
 \hline
$ J_3^{\mathbb{C}} $ &$SL(3,\mathbb{C})/U(3)$&$SU(3,3)/SU(3)^2 \times U(1) $ &$E_{6(2)}/SU(6)\times SU(2) $ \\
\hline $J_3^{\mathbb{H}}$ & $SU^*(6)/USp(6)$ &$SO^*(12) /U(6)$ &$E_{7(-5)}/SO(12)\times SU(2)$  \\ \hline
$J_3^{\mathbb{O}} $ & $E_{6(-26)}/F_4$ & $E_{7(-25)}/E_6 \times U(1)$ & $E_{8(-24)}/E_7 \times SU(2) $  \\ \hline

\end{tabular}

\end{small}
\end{center}

 Note that the exceptional $N=2$ supergravity has  $ E_{6(-26)}, E_{7(-25)}$ and $E_{8(-24)}$ as
  its global symmetry groups in five , four and three dimensions, respectively,
 whereas the maximally extended supergravity theory has the maximally split real forms $E_{6(6)}, E_{7(7)}$ and $E_{8(8)}$
 as its symmetry groups in the corresponding  dimensions.

 In addition to four simple Euclidean Jordan algebras $J_3^{\mathbb{A}}$ there exist an infinite family of
 nonsimple Jordan algebras of degree three, which are direct sums $ J = \mathbb{R} \oplus \Gamma(Q)$ of a one dimensional
 Jordan algebra $ \mathbb{R}$ with a Jordan algebra $\Gamma(Q)$
 associated with a quadratic form $Q$  \footnote{ The positivity of the kinetic terms of scalars and vectors requires the
 metric of the quadratic form to me Minkowskian .}. This family of $N=2$ MESGT's is called the generic Jordan family and
  their scalar manifolds in five dimensions are \[ \mathcal{M}_5(\mathbb{R} \oplus \Gamma(Q)) = \frac{SO(1,1) \times SO(n-1,1)}{SO(n)}
  \]
  and $SO(1,1) \times SO(n-1,1)$ are  the global symmetry groups of their Lagrangians.
The corresponding $N=2$ MESGT's in $d=4$ obtained by dimensional reduction have the scalar manifolds:
\[ \mathcal{M}_4 (\mathbb{R} \oplus \Gamma(Q)) = \frac{SO(2,1) \times SO(n,2)}{SO(2)\times SO(n) \times SO(2)} \]
By further reduction to three dimensions the scalar manifolds become
\[ \mathcal{M}_3 ((\mathbb{R} \oplus \Gamma(Q)) = \frac{ SO(n+2,4)}{ SO(n+2) \times SO(4)}
\]

 The term U-duality was introduced by Hull and Townsend since the discrete symmetry group
$E_{7(7)}(\mathbb{Z})$ of M/superstring theory toroidally compactified to $d=4$ {\it unifies} the T-duality group
$SO(6,6)(\mathbb{Z})$ with the S-duality group $SL(2,\mathbb{Z})$ in a simple group since
\[ SO(6,6) \times SL(2,\mathbb{R}) \subset E_{7(7)} \]
The analogous decomposition of the symmetry group $E_{7(-25)}$ of the exceptional supergravity in $d=4$ is
\[ SO(10,2) \times SL(2,\mathbb{R}) \subset E_{7(-25)} \]
with similar decompositions for the other magical supergravity theories. For the generic Jordan family of $N=2$
MESGT's there is no simple U-duality group that unifies the corresponding  S ( $SL(2,\mathbb{R})$) and T-duality
$( SO(n+2,2)) $ groups.

\section{U-duality Groups and  Entropy of BPS Black Holes in Supergravity Theories}

The entropy of BPS  black hole solutions in maximally extended supergravity as well as in matter coupled
supergravity theories are invariant under the corresponding U-duality groups. For example in $d=5$ , $N=8$
supergravity the entropy $S$  of a BPS black hole solution can be written in the form \cite{d5bh}

\begin{equation}
S= \alpha \sqrt{I_3} =\alpha  \sqrt{ d_{IJK} q^Iq^Jq^K}
\end{equation}
where $\alpha$ is some fixed constant and $I_3$ is the cubic invariant of $E_{6(6)}$ with the $q^I ,
I=0,2,...,26$ denoting the charges coupling to  27 vector fields of the theory. The BPS black hole solutions with
$I_3\neq 0$ preserve 1/8 supersymmetry \cite{fema} and the solutions with $I_3=0$ , but with $\frac{1}{3}
\partial_I I_3 = d_{IJK} q^Jq^K \neq 0$ preserve 1/4 supersymmetry. The BPS black hole  solutions with both $I_3=0$ and
$d_{IJK} q^Jq^K=0$ preserve 1/2 supersymmetry \cite{fema}. The orbits of the BPS black hole solutions of $N=8$
supergravity in $d=5$ under the action of its U-duality group $E_{6(6)}$ were classified in \cite{fegu}.

The entropy of the BPS black hole solutions of five dimensional $N=2$ MESGT's is given by the cubic form defined
by the constant tensor \cite{d5bh}
\[ S =\alpha \sqrt{\mathcal{N}}=  \alpha  \sqrt{C_{IJK}q^Iq^Jq^K} \]
For those $N=2$ MESGT theories defined Jordan algebras of degree three this cubic form is the norm form and the
global symmetry group $G$ is its invariance group as explained  in the previous section. The orbits of the BPS
black hole solutions of $5d$,  $N=2$ MESGT's defined by Jordan algebras under the actions of their U-duality
groups were classified in \cite{fegu}. This was achieved by associating with a given BPS solution with charges
$q^I$, $ (I=0,1,...n)$  an element \[ J=\sum_{I=0}^{n} e_Iq^I \]  of the Jordan algebra of degree three, where
$e_I$ form a basis of the Jordan algebra. This establishes a correspondence between the Jordan algebra and the
charge space of the BPS black hole solutions.

Similarly , the classification of the orbits of BPS black hole solutions of the $N=8$ supergravity in $d=5$ as
given in \cite{fegu} associates with a given BPS black hole solution with charges $q^I$ an element
$J=\sum_{I=0}^{26} e_Iq^I $ of the split exceptional Jordan algebra with basis elements $e_I$ \footnote{ Split
exceptional Jordan algebra $J_3^{\mathbb{O}_s}$ is realized in terms of hermitian $3\times 3$ matrices over the
split octonions $\mathbb{O}_s$ , which is not a division algebra. As a consequence  $J_3^{\mathbb{O}_s}$ is not a
Euclidean ( formally real) Jordan algebra in contrast to the real exceptional Jordan algebra $J_3^{\mathbb{O}}$
of  the exceptional $N=2$ theory, which is defined over the division algebra of real octonions $\mathbb{O}$. A
Jordan algebra $J$  is called Euclidean if for any pair of elements $ X,Y \in J$ the equation $X^2 +Y^2 =0 $
implies that $X=0$ and $Y=0$.}. The cubic invariant $I_3(q^I)$ is then simply given by the norm form
$\mathcal{N}$ of the split exceptional Jordan algebra. Invariance group $E_{6(6)}$ of this norm form ( known as
the reduced structure group in mathematics literature) is  the U-duality group of the maximal $N=8$ supergravity
theory in $d=5$.

In four dimensional magical $N=2$ MESGT's obtained by dimensional reduction from five dimensions as well as the
maximal $ 4d$,  $N=8$ supergravity the entropies of BPS black hole solutions are given by the quartic invariants
of their U-duality groups \cite{d4bh}

\begin{equation}
S= \beta \sqrt{I_4} =\beta \sqrt{d_{IJKL} q^Iq^Jq^Kq^L}
\end{equation}

where $d_{IJKL}$ are the completely symmetric tensors defined by the  Freudenthal-Kantor triple systems
associated with the corresponding simple Jordan algebras of degree three \cite{fegu} and $q^I$ now denote both
electric and magnetic charges. For the generic Jordan family of $N=2$ MESGT's in $d=4$ the quartic invariants are
defined by the completely symmetric tensors of the Freudenthal-Kantor triple systems associated with the groups
$SO(n+4,4)$. The orbits of the BPS black hole solutions of these theories under the action of the corresponding
U-duality groups were given in \cite{fegu}. For the maximal $N=8$ supergravity $q^I$ represent 28 electric and 28
magnetic charges. The number of supersymmetries preserved by the extremal black hole solutions of the $N=8$
theory depends on whether or not $I_4$ , $\partial_{J}I_4$ and $\partial_{J}
\partial_{K} I_4 $ vanish \cite{fema}.

\section{Generalized space-times  defined by Jordan  algebras}
\subsection{ Generalized Rotation, Lorentz and Conformal Groups}

In the previous sections we saw how Jordan algebras arise in a fundamental way within the framework of
supergravity theories. In this section I will review how Jordan algebras appear naturally within the framework of
attempts to generalize four dimensional Minkowski spacetime and its symmetry groups. The first proposal to use
Jordan algebras to define generalized spacetimes was made in the early days of spacetime supersymmetry while
searching for the  super analogs of the exceptional Lie algebras \cite{mg75}.

Now the twistor formalism in four-dimensional space-time $(d=4)$ leads naturally to the representation of four
vectors in terms of $2\times 2$ Hermitian matrices over the field of complex numbers ${\mathbb C}$.  In
particular, the coordinate four vectors $x_{\mu}$ can  be represented as :
\[ x=x_{\mu}\sigma^{\mu} \]
Since the Hermitian matrices over the field of complex numbers close under the symmetric anti-commutator product
one  can regard the coordinate vectors as elements of a Jordan algebra denoted as $J_2^{\mathbb C}$
\cite{mg75,mg80}.
  Then the rotation, Lorentz and conformal groups in
$d=4$ can be identified with the automorphism , reduced  structure and M\"{o}bius ( linear fractional) groups of
the Jordan algebra of
 $2\times 2$
complex Hermitian matrices $J_2^{\mathbb C}$ \cite{mg75,mg80}.
 The reduced structure group $Str_0(J)$ of a Jordan algebra $J$ is
 simply the
invariance group of its norm form $N(J)$. (The structure group
 $Str(J)= Str_0(J)\times SO(1,1)$
,on the other hand, is simply the invariance group of $N(J)$ up to an overall constant scale factor.)
Furthermore, this interpretation leads one naturally to define generalized space-times whose coordinates are
parameterized by the elements of  Jordan algebras  \cite{mg75}. The rotation $Rot(J)$, Lorentz $Lor(J)$ and
conformal $Con(J)$ groups of these generalized
 space-times are then identified with the automorphism $Aut(J)$,
 reduced structure $Str_0(J)$
and M\"{o}bius  M\"{o}(J) groups of the corresponding Jordan
 algebra \cite{mg75,mg80,mg91,mg92}.
  Denoting as  $J_{n}^{\mathbb A}$ the Jordan algebra of $n\times
n$   Hermitian matrices over the {\it division} algebra ${\mathbb A}$ and the Jordan algebra of Dirac gamma
matrices in $d$ ( Euclidean) dimensions as $\Gamma(d)$ one finds the following symmetry groups of  generalized
space-times defined by simple Euclidean ( formally real) Jordan algebras:
\begin{center}
\begin{tabular}{|c|c|c|c|}
 \hline
$J $& $Rotation(J)$ & $Lorentz(J)$ & $Conformal(J)$ \\
\hline
~&~&~&~\\
$J_{n}^{\mathbb R}$ & $SO(n)$ & $SL(n,{\mathbb R})$ & $Sp(2n,{\mathbb R})$\\
~&~&~&~\\
$J_{n}^{\mathbb C}$ & $SU(n) $ & $ SL(n,{\mathbb C})$ & $SU(n,n)$ \\
~&~&~&~\\
$J_{n}^{\mathbb H} $& $USp(2n)$ & $SU^{*}(2n)$ &$ SO^{*}(4n)$ \\
~&~&~&~\\
$J_{3}^{\mathbb O}$ &$ F_{4}$ & $E_{6(-26)}$ & $E_{7(-25)}$ \\
~&~&~&~ \\
$\Gamma(d)$ & $SO(d)$ & $SO(d,1)$ & $SO(d,2)$ \\
~&~&~&~ \\
\hline
\end{tabular}

\end{center}

The symbols ${\mathbb R}$, ${\mathbb C}$, ${\mathbb H}$, ${\mathbb O}$
 represent the four division
algebras.  For the Jordan algebras $J_n^{\mathbb A}$ the norm form is the determinental form ( or its
generalization to the quaternionic and octonionic matrices). For the Jordan algebra $\Gamma(d)$ generated by
 Dirac gamma matrices $\Gamma_{i} ~( i =1,2,...d)$

\begin{equation}
~\\~ \{\Gamma_i,\Gamma_j\} = \delta_{ij} {\mathbf 1};
~~~~~i,j,\ldots~=~ 1,2,\ldots,d \\
~\\
\end{equation}

the norm of a general element $x= x_0 {\mathbf 1} + x_i \Gamma_i$ of $\Gamma(d)$ is quadratic and given by \[
N(x) = x \bar{x}= x_0^2 -x_ix_i \] where $\bar{x}= x_0 {\mathbf 1} - x_i \Gamma_i $. Its automorphism, reduced
structure and M\"{o}bius groups are simply the rotation, Lorentz and conformal groups of $(d+1)$-dimensional
Minkowski spacetime. One finds the following special isomorphisms between the Jordan algebras of $2\times 2$
Hermitian matrices over the four division algebras and the Jordan algebras of gamma matrices:

\begin{equation}
~\\
J_{2}^{\mathbb R} \simeq \Gamma(2)~~~~;~~~~J_{2}^{\mathbb C} \simeq \Gamma(3) \\
~~~~;~~~~
J_{2}^{\mathbb H} \simeq \Gamma(5)~~~~;~~~~J_{2}^{\mathbb O} \simeq \Gamma(9) \\
~\\
\end{equation}

The Minkowski spacetimes they correspond to are precisely the critical dimensions for the existence of super
Yang-Mills theories as well as of the classical Green-Schwarz superstrings. These Jordan algebras are all
quadratic and their norm forms are precisely the quadratic invariants constructed using the Minkowski metric.

We should note two  remarkable facts about  the above table. First , the conformal groups of generalized
space-times defined by Euclidean ( formally real) Jordan algebras all admit positive energy unitary
representations  \footnote{ Similarly, the generalized conformal groups defined by
 Hermitian Jordan triple
systems all admit positive energy  unitary  representations \cite{mg92}. In fact the conformal groups of simple
Hermitian Jordan triple systems exhaust the list of simple noncompact groups that admit positive energy unitary
representations. They include the conformal groups of simple Euclidean Jordan algebra since the latter  form an
hermitian Jordan triple system under the Jordan triple product \cite{mg92}. }. Hence they can be given a  causal
structure with a unitary time evolution as in  four dimensional Minkowski space-time. Second is the fact that the
maximal compact subgroups of the generalized conformal groups of formally real Jordan algebras are simply the
compact forms of their structure groups (which are the products of their generalized Lorentz groups with
dilatations), whose importance will be explained in the next subsection.

\subsection{Covariant Quantum Fields over Generalized Spacetimes and the Positive Energy Unitary Representations of
Their Conformal Groups}

The Lie algebra $g$ of a non-compact group G that admits unitary lowest weight representations (ULWR) ( positive
energy )  has a 3-grading
 with respect to the Lie algebra $h$ of its maximal
compact subgroup H i.e

\begin{eqnarray}
g=g^{-1}\oplus g^0\oplus g^{+1}
\end{eqnarray}
where $g^0=h$ and we have the formal commutation relations
\begin{eqnarray}
{[}g^{(m)},g^{(n)}{]}\subseteq g^{(m+n)}\hspace{2.0cm}m,n=\mp 1,0\nonumber
\end{eqnarray}
and $g^{(m)}\equiv 0$ for $\vert m\vert >1.$ \vskip 0.3 cm

 In \cite{gs} the general  oscillator construction of
unitary lowest weight representations
 of non-compact groups  was given.
To construct the ULWR's of a noncompact group $G$ one  realizes its  generators  in terms of bilinears of bosonic
oscillators transforming in a certain representation of $H$. Then in the corresponding Fock space ${\cal F}$ of
all the oscillators  one chooses  a set of states $|\Omega >$, referred to as the "lowest weight vector" (lwv),
which transforms irreducibly under  $H$ and which
 are annihilated by the generators belonging to the
$g^{-1}$ space. Then by acting on $|\Omega >$ repeatedly with the generators belonging to the $g^{+1}$ space one
obtains an infinite set of states \[ |\Omega >,\ \ g^{+1} |\Omega >,\ \ g^{+1}g^{+1} |\Omega >,... \] that form
the basis  of an irreducible unitary lowest weight
 representation of $g$. (The
 irreducibility of the representation of $g$ follows from the
irreducibility of  the lwv $ |\Omega >$  under $h$. We shall refer to this basis as the compact "particle" basis.

 For example,  the conformal group of the Jordan algebra
$J_2^{\mathbb C}$ corresponding to the four dimensional Minkowski space
 is $SU(2,2)$  with a maximal compact subgroup
$SU(2)\times SU(2)\times U(1)$ which is simply the compact form of the structure group $SL(2,\mathbb C)\times
SO(1,1)$. In \cite{gmz2} it was explicitly shown how to go from the  compact $SU(2)\times SU(2) \times U(1)$
basis of the ULWR's of $SU(2,2)$ to the manifestly covariant  $SL(2,\mathbb C)\times SO(1,1)$ basis. The
transition from the compact  to the covariant basis corresponds simply to going from a compact "particle" basis
to a noncompact coherent state basis of the ULWR. The noncompact coherent states are labelled by the elements of
$J_2^{\mathbb C}$ i.e by the coordinates of four dimensional Minkowski space. One thereby establishes a
one-to-one correspondence between irreducible ULWR's of $SU(2,2)$ and the fields transforming irreducibly under
the Lorentz group $SL(2,{\mathbb C}) $ with a definite conformal dimension. Thus one can associate with
irreducible ULWR's of $SU(2,2)$  conformal fields transforming covariantly under the Lorentz group with a
definite conformal dimension.

Similarly, the conformal group $SO^*(8)$  of the Jordan algebra $J_2^{\mathbb H}$ parameterizing the six
dimensional Minkowski space has a maximal compact subgroup $U(4)$  which is the compact form of the structure
group $SU^*(4)\times SO(1,1)$. In \cite{mgst} it was shown explicitly how to go from the  compact $U(4)$ particle
basis of the ULWR's of $SO^*(8)$ to the non-compact  $SU^*(4) \times SO(1,1)$ coherent state basis. ( $SU^*(4)
\times SO(1,1)$  is simply the covering group of the Lorentz group in six dimensions times dilatations) . The
coherent states of the non-compact basis are again labelled by the elements of $J_2^{\mathbb H}$, i.e the
coordinates of 6d Minkowski space. Thus each irreducible ULWR of $SO^*(8)$ can be identified with a field
transforming covariantly under the Lorentz group $SU^*(4)$ with a definite conformal dimension.

The results obtained explicitly for the conformal groups of $J_2^{\mathbb C}$ and $J_2^{\mathbb H}$ extend to the
conformal groups of all formally real Jordan algebras and of Hermitian Jordan triple systems
\cite{mg2000,mg2005}. The general theory can be summarized as follows: Let $g$ be the Lie algebra of the
conformal group of a formally real Jordan algebra and $g^0$ the Lie algebra of its maximal compact subgroup. Then
$g$ has a three-graded decomposition with respect to $g^0$:
\begin{equation}
 g=g^- +g^0 + g^+  \nonumber
\end{equation}
where the grading is determined by the "conformal energy operator". Now let $n^0$ be the Lie algebra of the
structure group ( Lorentz group times dilatations ) of the Euclidean Jordan algebra ( or of a Hermitian Jordan
triple system). As stated above the Lie algebra  $g$ has a 3-graded decomposition with respect to $n^0$ as well:
\begin{equation}
  g=n^- + n^0 + n^+
\end{equation}
where the grading is defined by the generator of scale transformations. In the compact basis an irreducible ULWR
of $Conf(J)$ is uniquely determined by a lowest weight vector $| \Omega \rangle $ transforming irreducibly under
the maximal compact subgroup $K$ that is annihilated by the operators belonging to $g^-$
\begin{equation}
  g^-| \Omega \rangle =0
\end{equation}

As was done explicitly for the conformal groups in 4 and 6 dimensions \cite{gmz2,mgst} one can show that there
exists a rotation operator $W$ in the representation space with the property that the states  $W| \Omega \rangle
$ are annihilated by all the generators belonging to $n^-$
\begin{equation}
  n^-W| \Omega \rangle =0
\end{equation}
and  transform  in a finite dimensional non-unitary representation of the non-compact structure group. Remarkably
the transformation properties of $W| \Omega \rangle $ under the structure group coincide with the transformation
properties of $| \Omega \rangle $ under the maximal compact subgroup $K$. In particular, the conformal dimension
of the vector $W| \Omega \rangle $ is simply the  negative of the conformal energy of $| \Omega \rangle $. If one
chooses a basis $e_{\mu}$ for the Jordan algebra $J$ and denote the generators of generalized translations in the
space $n^+$ corresponding to $e_{\mu}$ as $P_{\mu}$, then the noncompact coherent states defined by the action of
generalized translations on $W| \Omega \rangle $
\begin{equation}
  |\Phi(x_{\mu} \rangle :=e^{i x^{\mu}P_{\mu}} W| \Omega \rangle
\end{equation}
form the covariant basis of the ULWR of the generalized conformal group $Con(J)$ \footnote{We should note that
the (super) coherent states associated with ULWR's of non-compact (super) groups introduced in \cite{bg} are
labelled by complex (super) "coordinates" in the compact basis. These (super) coordinates parametrize the (super)
hermitian symmetric space $G/H$. }  . The coherent states $|\Phi(x_{\mu} \rangle $ labelled by the coordinates
correspond to conformal fields transforming covariantly under the Lorentz group with a definite conformal
dimension. Since the state $W| \Omega \rangle $ is annihilated by the generators of special conformal
transformations $K_{\mu}$ belonging to the space $n^-$ this proves that the irreducible ULWR's are equivalent to
representations induced by finite dimensional irreps of the Lorentz group with a definite conformal dimension and
trivial special conformal transformation properties. This generalizes the well-known construction of the positive
energy representations of the four dimensional conformal group $SU(2,2)$ \cite{mack} to all generalized conformal
groups of formally real Jordan algebras and Hermitian Jordan triple systems. They are simply induced
representations with respect to the maximal parabolic subgroup $Str(J) \odot S_J$ where $ \odot$ denotes
semi-direct product and $S_J$ is the Abelian subgroup generated by generalized special conformal transformations.

We should perhaps note that the generalized Poincar\'{e} groups associated with the spacetimes defined by Jordan
algebras are of the form
\begin{equation}
  \mathcal{PG}(J) := Lor(J) \odot T_J
\end{equation}
where $T_J$ is the Abelian subgroup generated by generalized translations $P_{\mu}$.  For quadratic Jordan
algebras, $\Gamma(d)$ , $\mathcal{PG}(\Gamma(d)) $ is simply the Poincar\'{e} group in $d$ dimensional Minkowski
space. The group $\mathcal{PG}(\Gamma(d)) $ has a quadratic Casimir operator $M^2 = P_{\mu} P^{\mu}$ which is
simply
 the mass operator. For Jordan algebras $J$ of degree $n$ the
generalized Poincar\'{e} group $\mathcal{PG}(J)$ has a Casimir invariant of order $n$ constructed out of the
generalized translation generators $P_{\mu}$. For example for the real exceptional Jordan algebra $J_3^{\mathbb
O}$ the corresponding Casimir invariant is cubic and has the form
\begin{equation}
M^3 =  C_{\mu\nu\rho} P^{\mu}P^{\nu}P^{\rho}
\end{equation}
where $C_{\mu\nu\rho}$ is the symmetric invariant tensor of the generalized Lorentz group $E_{6(-26)}$ of
$J_3^{\mathbb O}$ ( $\mu, \nu, \rho,..=0,1,...26) $\footnote{ We should note that we are relabelling the indices
$I,J,..$  with Greek characters $\mu, \nu, ...$ so as to emphasize the fact that ,in this section,  we are
looking at  Jordan algebras from a spacetime point of view.}.

\section{ Conformal and quasi-conformal extensions  of U-duality groups in five and four dimensions}
As discussed above we can associate an element $J=\sum_{I=0}^{26} e_Iq^I $ of the real ( split) exceptional
Jordan algebra $J_3^{\mathbb{O}} ( J_3^{\mathbb{O}_s} )$ with a BPS black hole solution of the exceptional $N=2 (
N=8)$ supergravity theory with charges $q^I$ in five dimensions. Its entropy $S$ is given by the square root of
the cubic norm $\mathcal{N}(J)$ of $J$, whose invariance group is $E_{6(-26)}$ ( $E_{6(6)}$).

Acting on an element $J=\sum_{I=0}^{26} e_Iq^I $ of $J_3^{\mathbb{O}} ( J_3^{\mathbb{O}_s} )$ by its conformal
group $E_{7(-25)} ( E_{7(7)}) $ changes its norm and hence the corresponding entropy.  Thus one can regard
$E_{7(-25)} ( E_{7(7)}) $ as a spectrum generating symmetry in the charge space of BPS black hole solutions of
the exceptional $N=2$ ( $N=8$) supergravity in five dimensions \cite{gkn1,gkn2,gupa}. If one defines a distance
function between any two solutions with charges $q^I$ and $q'^{I}$ as the cubic norm of their difference
\[ d(q,q') \equiv \mathcal{N}_3(J-J') \] one finds that the light like separations are preserved under the
conformal action of $E_{7(-25)} ( E_{7(7)}) $ \cite{gkn1,gupa}. The explicit action of $E_{7(7)}$ and
$E_{7(-25)}$  on the corresponding 27 dimensional spaces are given in \cite{gkn1} and \cite{gupa} , respectively.
Let us review briefly the conformal action of $\EE$ given in \cite{gkn1}.  Lie algebra of $\EE$ has a 3-graded
decomposition
\[
\rep{133} = \rep{27}\oplus(\rep{78}\oplus\rep{1})\oplus\rep{\overline{27}}
\]
under its $\EEE \times \cD$ subgroup, where \cD represents the dilatation group $SO(1,1)$. Under its maximal
compact subalgebra \USp  Lie algebra \6[(6)] decomposes  as a symmetric tensor $\Gt^{ij}$ in the adjoint \rep{36}
of \USp and a fully antisymmetric symplectic traceless tensor $\Gt^{ijkl}$ transforming as the \rep{42} of \USp (
indices $1\le i,j,\ldots \le 8$ are  \USp indices). $\Gt^{ijkl}$ is traceless with respect to the real symplectic
metric $\gOd[ij]\!=\!-\gOd[ji]\!=\!-\gO[ij]$ (thus $\gOd[ik] \gO[kj]\!=\!\delta_i^j$). The symplectic metric is
used  to raise and lower indices, with the convention that this is always to be done from the left. The other
generators of conformal $\EE$ consist of  a dilatation generator $\Ht$, translation generators $\Et^{ij}$ and the
nonlinearly realized "special conformal" generators $\Ft^{ij}$, transforming as $\rep{27}$ and
$\rep{\overline{27}}$, respectively.

The fundamental \rep{27} of \6[(6)] on which  \7[(7)]) acts nonlinearly  can be represented as the  symplectic
traceless antisymmetric tensor $\Zt^{ij}$ transforming as \footnote{ Throughout we use the convention that
indices connected by a bracket are antisymmetrized with weight one.}

\[
\Gt^i{}_j(\Zt^{kl})     &=& 2\,\gd\oversym{ {}^k_j \Zt^{il}} \,, \non[1ex] \Gt^{ijkl}(\Zt^{mn})    &=&
\ft{1}{24}\eps[ijklmnpq]\Zt_{pq}\,,
 \lb{27-rep}
\]

where $ \Zt_{ij} \;:=\; \gOd[ik]\gOd[jl] \Zt^{kl} = (\Zt^{ij})\st\, \quad \text{ and } \quad \gOd[ij]\Zt^{ij}
\;=\; 0 $.
 The conjugate \rep{\overline{27}} representation transforms as

\[
\Gt^i{}_j(\Zbt^{kl})    &=& 2\,\gd\oversym{ {}^k_j \Zbt^{il}} \,, \non[1ex] \Gt^{ijkl}(\Zbt^{mn})   &=&
-\ft{1}{24}\eps[ijklmnpq]\Zbt_{pq}\,. \lb{27b-rep}
\]

 The cubic invariant of \6[(6)] in the \rep{27} is  given by
\[
  \cN_3 (\Zt) := \Zt^{ij} \Zt_{jk} \Zt^{kl} \gOd[il]\,.
\]

 The generators  $\Et^{ij}$ act as  translations on the space with coordinates $\Zt^{ij}$ as :
\[
  \Et^{ij} (\Zt^{kl}) = -\gO[i{[}k]\gO[l{]}j] -\ft18 \gO[ij]\gO[kl]
\]
and $\Ht$ by dilatations
\[
 \Ht (\Zt^{ij} ) = \Zt^{ij} \,.
\]
The  " special conformal generators" $\Ft^{ij}$ in the $\rep{\overline{27}}$ are realized nonlinearly:
\[
\Ft^{ij} (\Zt^{kl}) &:=& -2\,\Zt^{ij}(\Zt^{kl})
    + \gO[i{[}k]\gO[l{]}j] (\Zt^{mn}\Zt_{mn})
    +\ft18\, \gO[ij]\gO[kl] (\Zt^{mn}\Zt_{mn})\non[1ex]
&&{}+8\,\Zt\oversym{^{km}\Zt_{mn}\gO[n{[}i] \gO[j{]}l]}
    - \gO[kl](\Zt^{im}\gOd[mn]\Zt^{nj})
\lb{e7-nonlin-2}
\]
The norm form needed to define the $\EE$ invariant ``light cones'' is  constructed from the cubic invariant of
$\EEE$.  If we define the "distance" between $\Xt$ and $\Yt$ as  $\cN_3 (\Xt-\Yt)$ then it is manifestly
invariant under $\EEE$ and under the translations $\Et^{ij}$. Under $\Ht$ it transforms by a constant factor,
whereas under the action of $\Ft^{ij}$ one finds
\[
 \Ft^{ij} \Big(\cN_3(\Xt - \Yt) \Big) =
  (\Xt^{ij} + {\Yt}^{ij})  \cN(\Xt - \Yt)\,.
\lb{e7-nonlin-3}
\]
which proves that the light cone in $\Rn^{27}$ with base point $\Yt$ defined by
\[
\cN_3(\Xt - \Yt) = 0
\]
is indeed invariant under $\EE$.

The above formulas carry over in a straightforward manner to the  conformal realization of $E_{7(-25)}$ on a 27
dimensional space coordinatized by the real exceptional Jordan algebra $J_3^{\mathbb{O}}$. In this case the cubic
form is invariant under $E_{6(-26)}$ which has $USp(6,2)$ as a subgroup. The $USp(8)$ covariant formulas above
for $\EE$ are then replaced by $USp(6,2)$ covariant formulas \cite{gupa,gupa2}. For the generic Jordan family of
$N=2$ MESGT's in five dimensions the conformal extensions of U-duality groups $SO(1,1)\times SO(n-1,1)$ are the
groups
\[ SO(2,1) \times SO(n,2) \]
which act as spectrum generating symmetry groups in $d=5$. These conformal groups of generic $d=5$  MESGT's as
well as
 the conformal groups $\EE$ and $E_{7(-25)}$ acting on the 27 dimensional charge spaces of the $N=8$ and
the exceptional $N=2$ supergravity in five dimensions are isomorphic to the U-duality groups of the corresponding
four dimensional theories obtained by dimensional reduction. We should stress the obvious fact that the conformal
extensions of U-duality groups  act nonlinearly on the charge space of five dimensional theories , whereas the
isomorphic four dimensional U-duality groups  act linearly on the charge space spanned by both electric and
magnetic charges.

One may wander whether there exist "conformal extensions " of the four dimensional U-duality groups that act
nonlinearly  on the 4d  charge spaces as spectrum generating symmetry groups and   are isomorphic to the
U-duality groups of the corresponding three dimensional theories obtained by dimensional reduction. This question
was investigated in \cite{gkn1} and it was shown there that in the case of maximal supergravity,  even though
there is no conformal action of $E_{8(8)}$, which is the corresponding U-duality group in $d=3$,  it has a
quasi-conformal group action on a 57 dimensional space which is an extension of the 56 dimensional charge space
by an extra coordinate. For  BPS black hole solutions in $d=4$ this extra coordinate can be taken to be the
entropy \cite{gkn1}.

The realization of quasi-conformal action of $\E$ uses the 5-graded decomposition of its Lie algebra  with
respect to the Lie algebra of its $E_{7(7)}\times \cD$ subgroup
\[
\ba{c@{\,\,\oplus\,\,}c@{\,\,\oplus\,\,}c@{\,\,\oplus\,\,}c@{\,\,\oplus\,\,}c}
\fg^{-2} & \fg^{-1} & \fg^0 & \fg^{+1} &\fg^{+2}  \\[1ex]
\rep{1}  & \rep{56} & (\rep{133}\oplus\rep{1}) & \rep{56} & \rep{1} \ea \lb{e8-grading}
\]
with $\cD$ representing dilatations, whose generator together with grade $\pm 2$ elements generate an $SL(2,\Rn)$
subgroup. It turns out to be very convenient to work in a basis covariant with respect to the \SLR subgroup of
$\EE$ \cite{gkn1}. Let us denote the \SLR covariant generators belonging to the grade $-2,-1,0,1$ and $2$
subspaces in the above decomposition as follows:
\[
E \oplus \{E^{ij},E_{ij}\} \oplus \{G^{ijkl},\, G^i{}_j \, ;\, H\}
  \oplus \{F^{ij},F_{ij}\} \oplus F
\lb{e8-grading-gen}
\]
where $i,j,..=1,2...,8$ are now \SLR indices.

Consider now a  $57$-dimensional real vector space with coordinates
\[\cX:=(X^{ij},X_{ij},x) \] where $X^{ij}$ and $X_{ij}$ transform in the $28$ and $\tilde{28}$ of \SLR
and  $x$ is a singlet.   The generators of $\EE$ subalgebra act linearly on this space
\[
\begin{array}{rclrcl}
G^i{}_j(X^{kl}) &=& 2\,\gd\oversym{{}^k_j X^{il}} -\ft{1}{4}\gd^i_j X^{kl}\,, & G^{ijkl}(X^{mn})&=&
\ft{1}{24}\eps[ijklmnpq] X_{pq} \,,
\\[1ex]
G^i{}_j(X_{kl}) &=& -2\,\gd\undersym{{}^i_k X_{jl}}+\ft{1}{4}\gd^i_j X_{kl}\,, & G^{ijkl}(X_{mn})&=&
\gd^{[ij}_{mn} X^{kl]}_{\phantom{m}} \,,
\\[1ex]
G^i{}_j(x)    &=& 0 \,, & G^{ijkl}(x)   &=& 0 \,,
\end{array}\lb{e8-G}
\]
The generator $H$ of  dilatations  acts as
\[
\begin{array}{rclcrclcrcl}
H (X^{ij}) &=&  X^{ij} \,, &\quad& H (X_{ij}) &=&  X_{ij} \,, &\quad& H (x)    &=& 2\, x \,,
\end{array}\lb{e8-H}
\]
and the generator $E$  acts as translations on $x$:
\[
\begin{array}{rclcrclcrcl}
E(X^{ij}) &=& 0 \,, &\quad& E(X_{ij}) &=& 0 \,, &\quad& E(x) &=& 1 \,.
\end{array}\lb{e8-E}
\]
The grade $\pm 1$ generators act as
\[
\begin{array}{rclcrclcrcl}
E^{ij}(X^{kl}) &=& 0 \,, &\quad& E^{ij}(X_{kl}) &=& \gd{}^{ij}_{kl}  \,, &\quad&
E^{ij}(x)    &=& - X^{ij} \,, \\[1ex]
E_{ij}(X^{kl}) &=& \gd{}^{kl}_{ij} \,, &\quad& E_{ij}(X_{kl}) &=& 0 \,, &\quad& E_{ij}(x)    &=& X_{ij}\,.
\end{array}\lb{e8-Eij}
\]
The positive grade generators are realized nonlinearly. The generator $F$ acts as
\[
F(X^{ij}) &=&
 4 X\oversym{{}^{ik} X_{kl} X^{lj}}
 +X^{ij} X^{kl} X_{kl} \non
&&
 -\ft{1}{12}\eps[ijklmnpq] X_{kl} X_{mn} X_{pq}
 +X^{ij}\,x \,, \non[1ex]
F(X_{ij})  &=&
 -4 X\undersym{{}_{ik} X^{kl} X_{lj}}
 -X_{ij} X^{kl} X_{kl} \non
&&
 +\ft{1}{12}\epsd[ijklmnpq] X^{kl} X^{mn} X^{pq}
 +X_{ij}\,x \,, \non[1ex]
F(x) &=& 4\,\cI_4( X^{ij}, X_{ij})+x^2
\]
where $\cI_4$ is the quartic invariant of $\EE$
\[  \cI_4& \=
   X^{ij} X_{jk} X^{kl} X_{li}
 -\frac{1}{4} X^{ij} X_{ij} X^{kl} X_{kl}
+\ft{1}{96}\,\eps[ijklmnpq] X_{ij} X_{kl} X_{mn} X_{pq} \non
  &+\ft{1}{96}\,\epsd[ijklmnpq] X^{ij} X^{kl} X^{mn} X^{pq}
  \,\lb{e8-F}
\]
 The action of the remaining generators of $\E$ are as follows:
\[
F^{ij}(X^{kl}) &=&
 -4\, X\oversym{{}^{i[k} X^{l]j}}
 +\ft{1}{4}\,\eps[ijklmnpq] X_{mn} X_{pq} \,, \non[1ex]
F^{ij}(X_{kl}) &=&
 +8\,\gd\undersym{{}^{[i}_k X^{j]m}_{\phantom{m]n}} X_{ml}}
 +\gd{}^{ij}_{kl}\, X^{mn} X_{mn}   +2\, X^{ij} X_{kl}
 -\gd{}^{ij}_{kl}\,x \,, \non[1ex]
F_{ij}(X^{kl}) &=&
 -8\,\gd\oversym{{}_{[i}^k X_{j]m}^{\phantom{m}} X^{ml}}
 +\gd{}^{kl}_{ij}\, X^{mn} X_{mn}   -2\, X_{ij} X^{kl}
 -\gd{}^{kl}_{ij}\,x \,, \non[1ex]
F_{ij}(X_{kl}) &=&
 4\, X\undersym{{}_{ki} X_{jl}}
 -\ft{1}{4}\,\epsd[ijklmnpq] X^{mn} X^{pq} \,, \non[1ex]
F^{ij}(x) &=&
 4\, X\oversym{{}^{ik} X_{kl} X^{lj}} + X^{ij} X^{kl} X_{kl} \non
&&-\ft{1}{12}\,\eps[ijklmnpq]  X_{kl} X_{mn} X_{pq}
 + X^{ij}\,x \,, \non[1ex]
F_{ij}(x) &=&
 4\, X\undersym{{}_{ik} X^{kl} X_{lj}} + X_{ij} X^{kl} X_{kl} \non
&&-\ft{1}{12}\,\epsd[ijklmnpq]  X^{kl} X^{mn} X^{pq}
 - X_{ij}\,x  \,. \lb{e8-Fij}
\]

The above action of $\E$ was called quasiconformal in \cite{gkn1} since it leaves a certain norm invariant up to
an overall factor. Since the standard difference $ (\cX - \cY) $ of two vectors in the 57 dimensional space is
not invariant under "translations" generated by $(E^{ij},E_{ij})$, one defines a nonlinear difference that is
invariant under these  translations  as:
\[
\gd(\cX,\,\cY) \;:=\; (X^{ij}-Y^{ij},X_{ij}-Y_{ij}\;;\;x-y+\Sympl{X}{Y})\, = -\delta(\cY,\cX) \lb{difference}
\]
where $\Sympl{X}{Y}:= X^{ij}Y_{ij} -X_{ij}Y^{ij} $. One  defines  the norm of a vector $\cX$ in the 57
dimensional space as

\[
 \cN_4(\cX)\;\=\;\cN_4( X^{ij}, X_{ij}; x) \;:=\; \cI_4( X) - x^2\,, \lb{N4}
\]
Then the "distance" between any two vectors $\cX$ and $\cY$ defined as $ \cN_4(\gd(\cX,\cY)) $ is invariant under
$E_{7(7)}$ and translations generated by $E^{ij} , E_{ij}$ and $E$. Under the action of the remaining generators
of $\E$ one finds that
\[
F\Big(\cN_4(\gd(\cX,\cY))\Big) &=& 2\,(x+y)\,\cN_4(\gd(\cX,\cY)) \non F^{ij}\Big(\cN_4(\gd(\cX,\cY))\Big) &=&
  2\,(X^{ij}+Y^{ij})\,\cN_4(\gd(\cX,\cY)) \non
H\Big(\cN_4(\gd(\cX,\cY))\Big) &=& 4\,\cN_4(\gd(\cX,\cY)) \nn
\]
Therefore, for every $\cY\in \Rn^{57}$ the ``light cone'' with base point $\cY$, defined by the set of
$\cX\!\in\Rn^{57}$ satisfying
\[
\cN_4(\gd(\cX,\cY)) = 0\,,  \lb{E8lc}
\]
is preserved by the full $\E$ group.

The quasiconformal realization of the other real noncompact form $E_{8(-24)}$ with the maximal compact subgroup
$E_7 \times SU(2)$ is given in \cite{gupa,gupa2}. In going to $E_{8(-24)}$ the role played by the subgroup
$SL(8,\mathbb{R})$ of $E_{7(7)}$ is played by the subgroup $SU^*(8)$ of $E_{7(-25)}$. The quasiconformal groups
$E_{8(8)}$ and $E_{8(-24)}$ are isomorphic to the U-duality groups of the maximal $N=16$ supergravity and the
$N=4$ exceptional supergravity in three dimensions. Since their action changes the "norm" in the charge-entropy
space of the corresponding four dimensional theories they can be interpreted as spectrum generating symmetry
groups. The quasiconformal realizations of $\E$ and $E_{8(-24)}$ can be consistently truncated to quasiconformal
realizations of other exceptional subgroups.

For the generic Jordan family of $N=2$ MESGT's the quasi-conformal extensions of their U-duality groups $SO(2,1)
\times SO(n,2)$  in $d=4$ are $ SO(n+4,4) $, which are isomorphic to their U-duality groups in $d=3$.

 \section{ The minimal unitary representations of  U-duality groups as quasi-conformal groups   }

As we saw above the noncompact U-duality groups of four dimensional $N=2$ MESGT's defined by Jordan algebras of
degree three can also arise as spectrum generating conformal symmetry groups in five dimensions. These groups all
admit positive energy unitary representations which can be constructed using the oscillator method outlined in
previous sections. Here I would like to discuss the  unitary representations  of exceptional groups that one
obtains by quantization of their quasiconformal realizations \footnote{ We should note that both the
quasiconformal extensions of four dimensional U-duality groups and the  isomorphic three dimensional U-duality
groups act nonlinearly in their respective dimensions.}.  For $\E$ this was done in \cite{gkn2} and for
$E_{8(-24)}$ in \cite{gupa}. Remarkably, the quantization of the quasiconformal realizations of $\E$ and
$E_{8(-24)}$ yield their minimal unitary representations. The concept of a minimal unitary representation of a
non-compact group $G$ was first introduced by A. Joseph \cite{Joseph} and is defined as a unitary representation
on a Hilbert space of functions depending on the minimal number of coordinates for a given non-compact group.
 Here we shall summarize the results mainly for $\E$ and indicate how they
extend to $E_{8(-24)}$. By truncation one can obtain the minimal unitary realizations of smaller exceptional
groups as quasiconformal groups  as well as unitary realizations  of $\EE$ and other exceptional subgroups as
conformal groups \cite{gkn2,gupa,gupa2}.

Since the positive graded generators form an Heisenberg algebra one introduces 28 coordinates $X^{ij}$ and 28
momenta $P_{ij}\equiv X_{ij}$, and one extra real coordinate $y$ to represent the central term. By quantizing
\[
[X^{ij},P_{kl}] \;=\; \i
\]
we can realize the positive grade generators of $\E$ as
\[
E^{ij} \;:=\; y\,X^{ij}\,, \quad E_{ij} \;:=\; y\,P_{ij}\,, \quad E\;:=\;\ft12\,y^2.
\]
To realize the other generators of $\E$ one introduces a momentum conjugate to the coordinate $y$ representing
the central charge of the Heisenberg algebra:
\[
 [y,p] \;=\; \i
\]
Then the remaining generators are given by
\[ \lb{e7e8gen}
 H &:=& \ft{1}{2} (y\,p+p\,y) \,,\non[2ex] F^{ij} &:=& -p\,X^{ij} + 2\i y^{-1}\, \Com{X^{ij}}{I_4(X,P)}\non
       &=&  -4\,y^{-1} X\oversym{^{ik}P_{kl}X^{lj}}
            - \ft{1}{2}\,y^{-1}(X^{ij}P_{kl}X^{kl}+X^{kl}P_{kl}X^{ij})\non
       &&
         +\ft{1}{12} y^{-1} \epsilon^{ijklmnpq} P_{kl}P_{mn}P_{pq}
         -p\,X^{ij}\,,\non[2ex]
F_{ij} &:=& -p\,P_{ij} + 2\i y^{-1}\, \Com{P_{ij}}{I_4(X,P)}\non
       &=&  4\,y^{-1} P\undersym{_{ik}X^{kl}P_{lj}}
            +\ft12\,y^{-1}(P_{ij} X^{kl}P_{kl}+P_{kl}X^{kl} P_{ij})\non
       &&
          -\ft{1}{12} y^{-1} \epsilon_{ijklmnpq} X^{kl}X^{mn}X^{pq}
          -p\,P_{ij}\,,\non[2ex]
F      &:=& \ft12 p^2 + 2 y^{-2} I_4(X,P)    \non[2ex]
 G^i{}_j  &:=& 2\, X^{ik}P_{kj}
              +\ft{1}{4} X^{kl}P_{kl} \,\gd^i_j \,,\non[2ex]
G^{ijkl} &:=& -\ft{1}{2} X^{[ij} X^{kl]}
              +\ft{1}{48} \epsilon^{ijklmnpq} P_{mn} P_{pq} \,.
\]

The hermiticity of all generators is manifest. Here $I_4(X,P)$ is the fourth order differential operator

\[ \label{invgen}
I_4(X,P) &:=& -\ft12(X^{ij}P_{jk}X^{kl}P_{li}+P_{ij}X^{jk}P_{kl}X^{li})\non
         &&   +\ft18(X^{ij}P_{ij}X^{kl}P_{kl}+P_{ij}X^{ij}P_{kl}X^{kl})\non
         &&   -\ft{1}{96}\,\epsilon^{ijklmnpq} P_{ij}P_{kl}P_{mn}P_{pq} \non
         &&   -\ft{1}{96}\,\epsilon_{ijklmnpq} X^{ij}X^{kl}X^{mn}X^{pq}
               + \ft{547}{16} \,.
\]
and  represents the quartic invariant of $\EE$ because
\[ \label{quartic}
\Com{G^i{}_j}{I_4(X,P)} = \Com{G^{ijkl}}{I_4(X,P)} = 0 \,.
\]

The above unitary realization of $E_{8(8)}$ in terms of position and momentum operators ( Schr\"{o}edinger
picture) can be reformulated in terms of annihilation and  creation operators (oscillator realization) (
Bargman-Fock picture) \cite{gkn2}. The transition from the Schrodinger picture to the Bargmann-Fock picture
corresponds to going from the $SL(8,\mathbb{R})$ basis to the $SU(8)$ basis of $E_{7(7)}$ .

 The quadratic Casimir operator of $\E$ reduces to a number for the above realization and one can show that
 all the higher Casimir operators must also reduce to numbers as required  for an
irreducible unitary representation. Thus by exponentiating the above generators we obtain the minimal unitary
irreducible representation of $\E$ over the Hilbert space of square integrable complex functions in 29 variables.

The $SL(2,\mathbb{R})$ subgroup generated by the grade $\pm 2$ elements $E,F$ and the dilatation generator $H$
are precisely of the form that arises in conformal quantum mechanics \cite{AlFuFu76}. The quadratic Casimir of
this $SL(2,\mathbb{R})$ subgroup is simply
\[
C_2(SL(2,\mathbb{R})=  I_4 - \ft{3}{16} = \frac{g}{4} -\frac{3}{16} \] showing that the role played by the
coupling constant $g$ in conformal quantum mechanics is played by the quartic invariant $I_4$ of $E_{7(7)}$,
which is an $SL(2,\mathbb{R})$ singlet. The role of conformal quantum mechanics in the description of black hole
solutions of supergravity was discussed in \cite{CDKKTP98}.

In the minimal unitary realization of the other noncompact real form $E_{8(-24)}$ with the maximal compact
subgroup $E_{7} \times SU(2)$ given explicitly in \cite{gupa} the relevant 5-graded decomposition of its Lie
algebra $\mathfrak{e}_{8(-24)}$ is with respect to its subalgebra $\mathfrak{e}_{7(-25)} \oplus
\mathfrak{so}(1,1)$
\begin{equation}
 \mathfrak{e}_{8(-24)}  =
\begin{array}{ccccccccc}
   \mathbf{1} & \oplus & \mathbf{56} & \oplus & \left(\mathfrak{e}_{7(-25)} \oplus
\mathfrak{so}(1,1)  \right) & \oplus &
   \mathbf{56} & \oplus & \mathbf{1}
\end{array}
\end{equation}

The Schr\"{o}dinger picture for the minimal unitary representation of $E_{8(-24)}$ corresponds to working in the
$SU^*(8) $ basis of the $E_{7(-25)}$ subgroup.  The position and momentum operators  transform in the ${\bf 28}$
and ${\bf \tilde{28}}$ of this $SU^*(8)$ subgroup  and the above formulas for $\E$ carry over  to those of
$E_{8(-24)}$ with some subtle differences\cite{gupa}. The Bargmann-Fock picture for the minimal unitary
realization of $E_{8(-24)}$ in terms of annihilation and creation operators is obtained by going from the
$SU^*(8)$ basis to the $SU(6,2)$ basis of the $E_{7(-25)}$ subgroup of $E_{8(-24)}$.

One can obtain the minimal unitary realizations of certain subgroups of $E_{8(8)}$ and $E_{8(-24)}$ by truncating
their minimal realizations. However, we should stress that since the minimal realizations of $E_{8(8)}$
\cite{gkn2} and $E_{8(-24)}$ \cite{gupa} are nonlinear consistent truncations exist for only certain subgroups.
It turns out that the minimal unitary realizations of all lower rank noncompact exceptional groups can be
obtained from those of $\E$ and  $E_{8(-24)}$ \cite{gkn1,gupa}.  The relevant subalgebras of
$\mathfrak{e}_{8(-24)}$ and $\mathfrak{e}_{8(8)}$ are those that are realized as quasi-conformal algebras , i.e
those that have a 5-grading
\[
  \mathfrak{g}= \mathfrak{g}^{-2} \oplus \mathfrak{g}^{-1} \oplus \mathfrak{g}^{0} \oplus \mathfrak{g}^{+1} \oplus
  \mathfrak{g}^{+2} \nonumber
\]
such that $\mathfrak{g}^{\pm 2}$ subspaces are  one-dimensional and $\mathfrak{g}^{0}= \mathfrak{h} \oplus \Delta
$ where $\Delta$ is the generator that determines the 5-grading.. Hence they all
 have  an  $\mathfrak{sl}(2,   \mathbb{R})$  subalgebra  generated  by
 elements of  $\mathfrak{g}^{\pm 2}$  and the generator  $\Delta$.  For the truncated subalgebra,
  the quartic invariant $\mathcal{I}_4$ will now  be that  of a  subalgebra $\mathfrak{h}$ of
 $\mathfrak{e}_{7(-25)}$ or of $\mathfrak{e}_{7(7)}$. Furthermore, this subalgebra must act on the
 grade $\pm 1$ spaces via symplectic representation. Below  we give the
 main chain of such subalgebras \cite{gupa}
\begin{equation}
 \mathfrak{h} = \mathfrak{e}_{7(-25)} \supset \mathfrak{so}^\ast(12) \supset \mathfrak{su}(3,3)
      \supset \mathfrak{sp}(6, \mathbb{R}) \supset \oplus_{1}^3 \mathfrak{sp}( 2, \mathbb{R} )
\supset \mathfrak{sp}( 2, \mathbb{R} ) \supset \mathfrak{u}(1)
\end{equation}
Corresponding quasi-conformal subalgebras read as follows
\begin{equation}
  \mathfrak{g} = \mathfrak{e}_{8(-24)} \supset \mathfrak{e}_{7(-5)} \supset \mathfrak{e}_{6(2)} \supset
      \mathfrak{f}_{4(4)}
   \supset \mathfrak{so}(4,4)
   \supset \mathfrak{g}_{2(2)} \supset \mathfrak{su}(2,1)
\end{equation}
The corresponding chains for the other real form $\mathfrak{e}_{8(8)}$ are
\begin{equation}
\mathfrak{h}= \mathfrak{e}_{7(7)} \supset \mathfrak{so(6,6)} \supset \mathfrak{sl(6,\mathbb{R}} \supset
\mathfrak{sp}(6, \mathbb{R}) \supset \oplus_{1}^3 \mathfrak{sp}( 2, \mathbb{R} ) \supset \mathfrak{sp}( 2,
\mathbb{R} ) \supset \mathfrak{u}(1)
\end{equation}
\begin{equation}
 \mathfrak{g}= \mathfrak{e}_{8(8)} \supset \mathfrak{e}_{7(7)} \supset \mathfrak{e}_{6(6)} \supset
     \mathfrak{f}_{4(4)}
 \supset \mathfrak{so}(4,4)
 \supset \mathfrak{g}_{2(2)} \supset \mathfrak{su}(2,1)
\end{equation}

The minimal unitary realizations of $\mathfrak{e}_{8(8)}$ and of $\mathfrak{e}_{8(-24)}$ can also be consistently
truncated to unitary realizations of certain subalgebras that act as regular conformal algebras with a 3-grading.
For $\mathfrak{e}_{8(8)}$ we have the following  chain of consistent truncations to conformal subalgebras
$\mathfrak{conf}$:
\begin{equation}
\mathfrak{conf} = \mathfrak{e}_{7(7)} \supset \mathfrak{so(6,6)} \supset \mathfrak{sl(6,\mathbb{R}} \supset
\mathfrak{sp}(6, \mathbb{R}) \supset \oplus_{1}^3 \mathfrak{sp}( 2, \mathbb{R} ) \supset \mathfrak{sp}( 2,
\mathbb{R} )
\end{equation}

 The corresponding chain of consistent truncations to conformal subalgebras for
$\mathfrak{e}_{8(-24)}$ is
\begin{equation}
\mathfrak{conf}=\mathfrak{e}_{7(-25)} \supset \mathfrak{so}^\ast(12) \supset \mathfrak{su}(3,3)
      \supset \mathfrak{sp}(6, \mathbb{R}) \supset \oplus_{1}^3 \mathfrak{sp}( 2, \mathbb{R} )
\supset \mathfrak{sp}( 2, \mathbb{R} )
\end{equation}

\section{Concluding Remarks}
The minimal unitary realizations of $\E$ and $E_{8(-24)}$ and their subgroups given in \cite{gkn2,gupa} can be
extended to all noncompact groups and  supergroups \cite{gupa2} \footnote{For $SL(2,\mathbb{R})$ the minimal
realization reduces to the conformal realization.} and formulated in a unified manner.  This unified construction
follows closely the formalism of unified construction of  nonlinear superconformal and quasi-superconformal
algebras in two dimensions \cite{BinGun}. The realization of the distinguished $SL(2,\mathbb{R})$ subgroup
generated by the grade $\pm 2$ elements in the unified construction is always of the form that arises in
conformal or superconformal quantum mechanics. The coherent state formulation of the minimal unitary
representations of noncompact groups and supergroups analogous to those  of conformal groups is currently under
investigation.

As I stated at the beginning of my talk, the discrete subgroups
$E_{(11-d)(11-d)}(\mathbb{Z})$ of the U-duality groups of M-theory
toroidally compactified to d dimensions are expected to be
non-perturbative symmetries of M-theory \cite{huto}. One expects
similarly that certain discrete subgroups of the U-duality groups of
matter coupled supergravity theories obtained by other
compactifications of M/superstring theory to be non-perturbative
symmetries in the respective dimensions. The major goal of the work
summarized in this talk is to find the unitary realizations of these
discrete subgroups and understand how the spectra of compactified
M/Superstring theory  fit into these representations.

 \vspace{1cm}

 {\bf Acknowledgement:} I would like to thank the
organizers of the XIXth Max Born Symposium in Wroclaw
 for their kind hospitality. The work presented here was done mostly in collaboration with Kilian Koepsell, Hermann Nicolai and
 Olexander Pavlyk, which I would like to acknowledge with pleasure. This work was supported in part by the
    National Science Foundation under grant number PHY-0245337. Any opinions, findings and conclusions or
     recomendations expressed in this material  are those of the author and do not necessarily
reflect the views of  the National Science Foundation.


\end{document}